\documentclass[aps,prd,twocolumn,superscriptaddress,preprintnumbers,nofootinbib]{revtex4}

\usepackage{epsfig}
\usepackage{amssymb}
\usepackage{amsmath}
\usepackage{amsfonts}
\usepackage{slashed}
\usepackage{hyperref,xcolor}

\newcommand{\as}{\alpha_s}
\newcommand{\Ord}{{\cal O}}
\newcommand{\met}{\slashed{E}_t}
\newcommand{\sss}{\scriptscriptstyle\rm}
\newcommand{\muf}{\mu_{\sss F}}
\newcommand{\mur}{\mu_{\sss R}}
\let\originalleft\left
\let\originalright\right
\renewcommand{\left}{\mathopen{}\mathclose\bgroup\originalleft}
\renewcommand{\right}{\aftergroup\egroup\originalright}
\newcommand{\be}{\begin{equation}}
\newcommand{\ee}{\end{equation}}

\newcommand{\lp}{\left(}
\newcommand{\rp}{\right)}

\begin{document}

\title{Signal-background interference effects for $gg \to H \to W^+ W^-$ beyond leading order}

\author{Marco Bonvini}
\affiliation{Deutsches Elektronen-Synchroton, DESY, Notkestra{\ss}e 85, D-22603 Hamburg, Germany}
\author{Fabrizio Caola}
\affiliation{Department of Physics and Astronomy, Johns Hopkins 
University, Baltimore, USA}
\author{Stefano Forte}
\affiliation{Dipartimento di Fisica, Universit\`a di Milano and INFN, Sezione di Milano, Via Celoria 16, I-20133 Milano, Italy}
\author{Kirill Melnikov}
\affiliation{Department of Physics and Astronomy, Johns Hopkins 
University, Baltimore, USA}
\author{Giovanni Ridolfi}
\affiliation{Dipartimento di Fisica, Universit\`a di Genova and INFN, Sezione di Genova, Via Dodecaneso 33, I-16146 Genova, Italy}

\preprint{DESY 13-059, IFUM-1011-FT}

\begin{abstract}
\vskip0.2cm
We study the effect of QCD corrections to the $gg\to H\to W^+ W^-$ 
signal-background interference at the LHC for a heavy  Higgs boson.
We construct a soft-collinear approximation to the
NLO and NNLO corrections for the background
process, which is exactly known only at LO. We 
estimate  its accuracy by constructing and
comparing the same approximation to the
exact result for the signal process, which is known up to NNLO, and we
conclude that we can describe the signal-background interference
to better than $\Ord(10\%$) accuracy.
We show that our 
result implies
that, in practice,
a fairly good  approximation to higher-order QCD corrections 
to  the interference  may also be obtained by 
rescaling the known  LO result by a $K$-factor computed using
the signal process.
\end{abstract}

\maketitle

\section{Introduction}
Search for  the Higgs boson  at the LHC has  been 
a remarkable success so far. Indeed, both  the ATLAS and CMS 
collaborations have announced the discovery of a new 
boson, whose properties  are  
compatible with that of the Standard Model  Higgs particle, 
with mass $m_h\approx 125$~GeV. Both collaborations also  excluded 
additional Higgs-like bosons in a large mass range 
$m_h\lesssim 600$~GeV~\cite{discovery-atlas,discovery-cms}.
The interpretation of the excesses observed in various production 
and decay channels, as 
originating from a single spin-zero particle,  
was made possible by detailed 
theoretical predictions for the Higgs boson production and
decay rates, see Ref.~\cite{hwg} for an overview. 

However, these experimental results do not imply 
that there are no additional Higgs-like bosons 
with masses $600~{\rm GeV} \lesssim m_h\lesssim 1~{\rm TeV}$. In fact,
the search for such particles is well underway~\cite{Chatrchyan:2013yoa}.
In the Standard Model, as the Higgs boson becomes heavier, 
its total decay width grows rapidly  $\Gamma_h \sim m_h^3$ thanks 
to contributions of the longitudinal electroweak bosons:
for $m_h\sim 600$~GeV, the width is close to $120~{\rm GeV}$. 
Since the finite-width effects change the distribution of the invariant 
masses of the decay products of the Higgs boson, 
their understanding is important for developing experimental 
search strategies.  

There are two finite width effects that influence the Higgs boson 
lineshape. First, the Higgs propagator must assume the Breit-Wigner 
form in the resonant regime $1/(s-m_h^2) \to 1/(s-m_h^2 + i m_h \Gamma_h)$.
While this modification is literally correct for a light (and therefore narrow) Higgs boson, 
for a heavy Higgs, it must be modified; the proper way to do this was 
subject to a significant discussion in recent literature, see 
Refs.~\cite{Goria:2011wa,Franzosi:2012nk} and
references therein.  The second 
effect is the interference with the background. Note that, in principle, 
the two effects are not completely independent of each other since 
modifications of the Breit-Wigner form for the propagator change the 
very definition of the ``background'' in the resonance region, but 
discussion of these subtleties is beyond the scope of this paper. 

Our goal is to consider the interference of the 
signal process $gg  \to H \to W^+W^-$ and the background process 
$gg \to W^+ W^-$ for a {\it heavy} Higgs boson\footnote{For the light 
$m_h = 125~{\rm GeV}$ Higgs boson the interference is negligible if 
proper signal-selection criteria are applied~\cite{keith,nikolas}.}.
This interference was first computed at leading order 
 in Refs.~\cite{binoth,keith}. Although the $gg \to W^+W^-$ 
amplitude appears at one loop, 
it is enhanced at the LHC by the large gluon flux, making 
the interference effects non-negligible. 
An obvious shortcoming of Refs.~\cite{binoth,keith}
is that their  analysis of the interference is performed at leading order 
in perturbative QCD as far as the Higgs boson signal is concerned. 
This is unfortunate since, for the Higgs boson signal, higher order QCD 
corrections are extremely important, as they enhance the total
rate by more than a factor two 
\cite{Harlander:2002wh,kb,Ravindran:2003um}. 
It is therefore interesting to explore their impact
on the signal-background interference.

Such an endeavor, however, is highly non-trivial. Indeed,  
a full NLO and NNLO QCD calculation  of  background amplitudes 
requires  evaluation of two- and three-loop $2 \to 2$ Feynman 
diagrams which is beyond the reach of the current computational technology. 
On the other hand, it is well-known~\cite{Kramer:1996iq}
that for the Higgs boson signal  
a large fraction of radiative corrections is  captured
by the soft-collinear approximation. Since this approximation 
should be particularly suitable for the description of a {\it heavy} 
Higgs boson,  we construct 
a soft-collinear approximation for the entire $gg \to W^+W^-$ amplitude 
that includes both the signal and the background and study the impact 
of these corrections on the interference. 

This paper is organized as follows. In Section~\ref{setup} 
we sketch the construction of the soft-collinear approximation.
In Section~\ref{numres} we present numerical results. We conclude 
in Section~\ref{conc}.

\section{Setup}\label{setup}
We begin by   describing  the setup 
of our computation. We are interested
in higher order QCD corrections to the interference between the signal process
$gg\to H\to W^+ W^-$ and the pure QCD background $gg\to W^+ W^-$.
We compute these
corrections in the soft gluon approximation, which is known to describe
the full NLO and NNLO Higgs cross section to very good accuracy. 
We will numerically assess the 
accuracy of our approximation in Sec.~\ref{numres} by comparing it 
with known NLO and NNLO
results for the signal process. 

The cross section for the production of a $W^+W^-$ pair with invariant mass 
$Q^2$, fully differential in the kinematics variables of the two $W$'s,
is given by
\begin{multline}\label{eq:rapdist}
{\rm d}\sigma\lp\tau,y,\{\theta_i\},Q^2\rp =
\int {\rm d} x_1 {\rm d} x_2 {\rm d}z\,
f_g(x_1,\muf) f_g(x_2,\muf)\\
\times \delta(\tau-x_1x_2z)\, 
{\rm d}\hat \sigma\lp z,\hat y,
\{\hat\theta_i\},\as,\frac{Q^2}{\mur^2},\frac{Q^2}{\muf^2}\rp
\end{multline}
where $f_g$ is the gluon distribution, and
${\rm d}\hat \sigma$ is 
the differential partonic cross section for the process 
\begin{equation}
g(p_1) + g(p_2)\to W^+(p_{W^+})+ W^-(p_{W^-})+X,
\end{equation}
with 
$(p_{W^+}+p_{W^-})^2=Q^2$; $\muf$ and $\mur$ are 
the factorization and the renormalization scales,
 $\as = \as(\mur)$ is the strong coupling 
constant at the scale $\mur$, $\tau\equiv Q^2/s$.
We denote by $y$ the rapidity of the $W$ pair,
and by $\{\theta_i\}$ a generic set of variables describing the kinematics
of the decay products of the $W^+W^-$ system in the hadronic
center-of-mass frame; they are related to the corresponding
variables $\hat y,\{\hat\theta_i\}$ in the partonic center-of-mass frames
by a boost with rapidity $y_{\rm cm}=\frac{1}{2}\ln\frac{x_1}{x_2}$, and
thus the $\hat\theta_i$ are functions of $\{\theta_i\}, x_1,x_2$ and $z$.

In the soft ($z\to 1$) limit, the rapidity
distribution of the $W^+W^-$ pair is entirely determined by
the inclusive
cross section~\cite{Bolzoni:2006ky,Becher:2007ty,Bonvini:2010tp}, up
to corrections suppressed by 
powers of $(1-z)$, and the partonic cross section in
Eq.~\eqref{eq:rapdist} takes the form 
\begin{multline}
{\rm d}\hat
\sigma\lp z,\hat y,
\{\hat\theta_i\},\as,\frac{Q^2}{\mur^2},\frac{Q^2}{\muf^2}\rp\\
={\rm d}\hat \sigma^{(0)}(\{\hat\theta_i\},\as)
z\, G\lp z,\as,\frac{Q^2}{\mur^2},\frac{Q^2}{\muf^2}\rp,
\end{multline}
where ${\rm d}\hat \sigma^{(0)}(\{\hat\theta_i\},\as)\delta(1-z)$ is the leading order partonic cross
section, and $G\lp z,\as,Q^2/\mur^2,Q^2/\muf^2\rp$
is the inclusive coefficient function computed in the soft limit,
i.e.\ (up to the explicit $z$ factor) the inclusive partonic cross section normalized to the leading order in
such a way that $G(z,\as)=\delta(1-z)+\Ord(\as)$.

In the same limit, the momenta of the $W$ bosons in the partonic center-of-mass frame are given by 
\be\label{wmom}
\hat p_{W^{\pm}} = \frac{\sqrt{Q^2}}{2}\lp1,\pm \beta \sin\hat\theta,0,\pm\beta\cos\hat\theta\rp
\ee
with $\hat\theta$ the $W$ boson scattering angle in the partonic center-of-mass frame,
and $\beta = \sqrt{1-4m_W^2/Q^2}$ (for simplicity, we have
assumed that the $W$-bosons are on-shell, but we will not make this assumption in the
sequel). The kinematics of the process in the soft limit is therefore the same
as the leading order kinematics, except that the total energy squared is rescaled by
a factor $z$. 

The boost that relates the partonic and hadronic center-of-mass frames
is fixed by taking for  the momenta of the colliding gluons
 either $p_1 = z x_1  P_1,\; p_2 = x_2 P_2$ or $p_1 = x_1 P_1,\;
 p_2 = z x_2  P_2$, where $P_{1,2}$ are four-momenta of the colliding
protons~\cite{Becher:2007ty}. Alternatively, one may also take as
momenta of the colliding gluons 
$p_1 = \sqrt{z} x_1  P_1,\; p_2 = \sqrt{z} x_2
P_2$~\cite{Bolzoni:2006ky}. These two choices coincide in the soft
limit up to terms suppressed by two powers of
$(1-z)$~\cite{Bonvini:2010tp} and, in fact, give very similar
results for observables considered in this paper. We will make the first choice at NLO,
where it is actually exact, while at NNLO we will take the average of
the results obtained with either choice cases.

We now turn to  the explicit form  of the coefficient function 
$G( z,\as,Q^2/\mur^2,Q^2/\muf^2)$, which contains the core of our
soft-collinear approximation.  We first sketch 
the important features of the soft gluon approximation and its 
modifications by focusing on the next-to-leading order. 
Further details on this, including required modifications 
at NNLO, can be found in Refs.~\cite{deflorian,higgs3}.

Working to  NLO accuracy and in the soft limit 
and neglecting all non-singular
terms, we  write the function $G$ as
(we suppress explicit scale dependence for simplicity)
\begin{multline}\label{defl_LO}
G(z,\as) = \delta(1-z)\\
+ \frac{\as}{2\pi} \left[
8 C_A {\cal D}_1(z) +
\lp \frac{2\pi^2}{3}C_A+ c_1\rp \delta(1-z)
\right] 
\end{multline}
where ${\cal D}_i(z) = \left[\ln^i(1-z)/(1-z)\right]_+$ and $c_1$ 
is the ratio of the infrared regulated higher-order 
virtual contributions to the  cross section and the leading order 
cross section for $gg \to W^+W^-$, 
see~\cite{deflorian} for its proper definition.\footnote{Because 
we consider here the $2 \to 2$ scattering process, $c_1$ does 
depend on the scattering angle.  We assume that this dependence 
is mild and systematically ignore it in this paper. Partial justification 
for this assumption is given below.}
For our purposes, the important feature of this formula is that non
universal NLO corrections for the process $gg\to WW$ only enter
through the coefficient $c_1$. This is because only emissions from external
gluon lines in each diagram contribute to the amplitude in the soft limit.
For the signal-only process $gg\to
H\to WW$, $c_1$ is known both in the infinite
$m_t$~\cite{nlo_dsz_pl,nlo_dawson} approximation and for finite
$m_t$~\cite{nlo_dsz}.  The determination of $c_1$ for the interference
would require the evaluation of complicated $gg \to W^+W^-$ amplitudes
which is beyond existing technical capabilities.

However,  we note that
the value of $c_1$ can be obtained without any computation in the 
kinematic limit 
$ 4 m_W^2 \ll Q^2 \ll 4 m_t^2$, 
$m_b \sim m_t$. 
In this limit, the interference is dominated by the contribution of  
longitudinally polarized $W$ bosons, which can be obtained from QCD 
corrections to the production of two neutral scalars $gg \to HH$ 
in the heavy top mass limit \cite{Dawson:1998py}.  Since both the box contribution for $gg \to HH$ 
and the triangle contribution for $gg \to H$ are described by the same effective Lagrangian, 
the virtual QCD corrections should be identical in the two cases.
Although the assumptions $Q^2\ll 4 m_t^2$, $m_t \sim m_b$ are not really
justified, we take
the value for $c_1$  that is obtained in 
 that limit as a reference value, and estimate the sensitivity of the 
final result to its variations. 

The soft approximation of
Eq.~\eqref{defl_LO} is of course only defined up to subleading terms. 
An optimal choice of subleading terms can be found~\cite{higgs3}
by using a combination of analiticity arguments in Mellin space, and
information on universal subleading terms in the $z\to1$ limit,
arising partly from the exact soft-gluon kinematics~\cite{Bonvini:2010tp}
and partly from universal collinear splitting kernels~\cite{Kramer:1996iq,Catani:2001ic}.
A discussion of this optimal soft approximation is beyond the scope 
of this paper, and we refer to Ref.~\cite{higgs3} for a full discussion. 
Here, we  note that the best approximation proposed in \cite{higgs3}
(called soft$_2$ there)
effectively amounts to performing in Eq.~\eqref{defl_LO} the replacement
\begin{align}
{\cal D}_i(z) &\to {\cal D}_i(z) + \delta {\cal D}_i(z),\nonumber\\
\delta {\cal D}_i(z) &= (2-3 z + 2 z^2) \frac{\ln^i\frac{1-z}{\sqrt z}}{1-z} -
\frac{\ln^i(1-z)}{1-z},
\label{sc}
\end{align}
where $\delta {\cal D}_i(z)$ is an ordinary function (not a distribution).
In what follows, we will call the approximation based on Eq.~\eqref{defl_LO}
with such replacement a ``soft-collinear'' approximation.
We will quantify the impact
of subleading effects by comparing this improved soft-collinear
approximation to a purely soft result.

At higher orders the soft approximation Eq.~\eqref{defl_LO} 
is also known: see e.g.\ Eq.~(79) in~\cite{deflorian}. 
We  improve it analogously  to Eq.~\eqref{sc}, see
Ref.~\cite{higgs3} for details.
This soft-collinear approximation is the basis for the NLO and NNLO numerical 
results for the signal and the interference that we discuss in the next Section.

\section{Numerical results}
\label{numres}

We consider the process $gg\to W^+(e^+\nu) W^-(e^-\bar\nu)$ at the LHC
for two values of the center-of-mass energy: $\sqrt s=8$~TeV and
$\sqrt s=13$~TeV.  We take the Higgs mass to be $m_h=600$~GeV, and its 
total decay width to be
$\Gamma_h=122.5$~GeV~\cite{hdecay}.  All numerical results presented
below are obtained with a fixed-width Breit-Wigner function.  
We have checked that use of the running-width
in the Breit-Wigner propagator~\cite{Dittmaier:2012vm} leads to
results for the signal and interferences that differ by an amount that
is below our accuracy goal, and we expect that same is likely to be
the case for a full treatment of finite-width
effects~\cite{Goria:2011wa,Franzosi:2012nk}.  Moreover, we have found
that the QCD radiative corrections are  insensitive to the
propagator, to the accuracy we work to.  We let both the $W$-bosons
decay leptonically and reconstruct all kinematic variables from
the charged lepton and neutrino momenta.  We take the $W$ total width to be
$\Gamma_W=2.11$~GeV and heavy quark masses $m_t=172.5$~GeV and $m_b=4.4$~GeV.

We use the NNPDF2.3 PDF set~\cite{Ball:2012cx} at NLO 
and NNLO, with $\as(m_Z)=0.118$.
Throughout this paper, we 
set the renormalization and factorization scales equal to the 
Higgs boson mass $\mur=\muf=m_h$.
In constructing our soft-collinear approximation, we retain the exact
$m_t$ and $m_b$ dependence where available. For example, we use the
exact value of $c_1$, Eq.~\eqref{defl_LO}, for the signal process,
while for the analogous $\Ord(\as^2)$ coefficient $c_2$ we use the
value computed in the infinite $m_t$ (point-like) approximation.  Note
that with this choice, all logarithmic terms at NNLO have the
exact $m_t$ and $m_b$ dependence, while the coefficient of
the $\delta(1-z)$ term is only
approximate.  As mentioned in Sect.~\ref{setup}, for the interference
we take the result in the $m_W^2\ll Q^2 \ll m_t^2$, $m_b\sim m_t$
limit as our reference value.

\begin{table}[t]
\centering
\begin{ruledtabular}
\begin{tabular}{cllcll}
& \multicolumn{2}{c}{$\sqrt s=8$~TeV} && \multicolumn{2}{c}{$\sqrt s=13$~TeV}\\
\cline{2-3}\cline{5-6}
& NLO & NNLO && NLO & NNLO \\
\hline
exact          & 2.150 & 2.78  && 2.074 & 2.67  \\
soft-collinear & 2.187 & 2.820 && 2.127 & 2.730 \\ 
$N$-soft       & 2.135 & 2.700 && 2.073 & 2.607 \\
\end{tabular}
\end{ruledtabular}
\caption{$K$-factors for the inclusive Higgs-only cross section in the narrow
width approximation, with $m_h=600$~GeV, computed using the exact
theory, our best soft-collinear approximation, and an
unimproved soft approximation  (see text for details). 
The (N)NLO result is computed using (N)NLO
PDFs, while the reference LO cross section is always computed with NLO PDFs.
Numerical results are obtained using the code \cite{higgscode}.}
\label{tab:inclKfactors}
\end{table}
To assess the quality of the soft-collinear approximation, we first
test it against the signal-only $gg\to H$ process at NLO and NNLO.
Results are shown in Tab.~\ref{tab:inclKfactors} for two values of the
collider energy. The $K$-factors computed (without
including the Higgs decay) using the exact theory\footnote{At NNLO, an
  exact result valid for large Higgs masses is not currently
  available. For our result, we use the exact result at NLO~\cite{nlo_dsz_pl} plus the
  point-like result at $O(\as^2)$, improving it with those $m_t$, $m_b$
  dependent terms which are fully determined by lower orders (which
  include all soft-collinear terms).  We have checked that the result
  obtained in this way is stable upon variation of small-$z$ terms up to
  the accuracy shown in Table~\ref{tab:inclKfactors}, which is a
  consequence of the dominance of soft-collinear terms for a heavy
  Higgs boson at the LHC~\cite{Bonvini:2012an}.} are compared to those obtained with
our soft-collinear approximation, or with the so-called $N$-soft
approximation, defined in Ref.~\cite{higgs3}. The latter amounts to
approximating the partonic cross section with the inverse Mellin
transform of a pure $N$-space soft approximation, in which only powers
of $\ln N$ and constant terms are kept.  

Both approximations reproduce
the exact result to $\Ord(3\%)$ or better in all configurations. At
$\sqrt{s}=8$~TeV, where the soft-collinear terms are expected to
dominate~\cite{Bonvini:2012an}, our soft-collinear approximation
reproduces the exact result to better than $\Ord(2\%)$, while at
higher energy, $\sqrt{s}=13$~TeV, the agreement deteriorates slightly,
because non-soft terms become relatively more important.  However,
whereas at NNLO the soft-collinear approximation is  more accurate than
the $N$-soft, at NLO the opposite happens. 
This occurs because numerically the $N$-soft approximation
happens to be closer to the exact result
than our improved soft-collinear one in the small-$N$ limit.
Since the small-$N$ limit is beyond the region of applicability
for both of these approximations, we consider this feature
to be accidental but  note that one can improve both of  these
approximations by matching them to the correct small-$N$ limit~\cite{simone}.
In what follows we use the soft-collinear approximation as
the default and take the spread of values between the soft-collinear and the
$N$-soft approximations as an estimate of the uncertainty due to
deficiencies of these approximations in the small-$N$ region.

We have also checked the reliability of
our approximation for differential distributions when
decays are included.  Indeed, at NLO accuracy, we find that our
approximate results for the lepton $p_t$ and rapidity distributions
and for the lepton invariant mass $m_{ll}$ distribution are in good
agreement with the full result obtained from MCFM
\cite{mcfm}. 

\begin{table}[t]
\centering
\begin{ruledtabular}
\begin{tabular}{clllclll}
& \multicolumn{3}{c}{$\sqrt s=8$~TeV} && \multicolumn{3}{c}{$\sqrt s=13$~TeV}\\
\cline{2-4}\cline{6-8}
& LO & NLO & NNLO && LO &NLO &NNLO\\
\hline
$\sigma_H$ & 0.909 & 1.99(5) & 2.6(1) && 3.77 & 8.1(2) & 10.3(5)\\
$\sigma_{Hi}$ & 1.188 & 2.6(1) & 3.4(3) && 4.56 & 9.7(4) & 12.5(9) \\
$\sigma_H/\sigma_H^{\rm LO}$ & --- &  2.19(5) & 2.8(1) && --- &  2.14(5) & \phantom{0}2.7(1) \\
$\sigma_{Hi}/\sigma_{Hi}^{\rm LO}$ & --- & 2.2(1) & 2.9(2) && --- & 2.13(9) &  \phantom{0}2.8(2) \\
\end{tabular}
\end{ruledtabular}
\caption{Results (in fb) for the Higgs-only cross section $\sigma_H$
  and the signal+interference cross section $\sigma_{Hi}$, with
  $m_h=600$~GeV. No cuts on the final state applied.
  The errors represent the uncertainty on the soft-collinear approximation and on 
  the unknown background coefficients,
  estimated as explained in the text.}
\label{restable}
\end{table}

Having assessed the accuracy of our approximation, we can now apply it
to study higher order corrections to the signal-background
interference.  As explained in the previous Section, we need the exact
leading order prediction for the interference.  We extract it
from~Ref.~\cite{keith}, as implemented in MCFM. For the Higgs boson
signal, we use the exact expression obtained as discussed above.
For the background, we include the contributions of all the three
quark generations, see~\cite{keith} for details. We also need the
infrared-regulated virtual cross section $c_1$, and the analogous NNLO
coefficient $c_2$.  As already mentioned, we take the signal values
for these coefficients $\bar c_{1,2}$ as a reference, and study the
impact of virtual corrections on the interference by varying $c_{1,2}$
in the range $-5 \bar c_{1,2} < c_{1,2} < 5 \bar c_{1,2}$.

We first discuss the impact of QCD corrections on the inclusive cross
section.  Following Ref.~\cite{keith}, we compare the signal-only
cross section $\sigma_H$ with the background-subtracted cross section
$\sigma_{Hi}\equiv \sigma_{gg\to WW} - \sigma_{gg\to WW}|_{\rm bg~
  only}$, which includes interference effects.  We report our results
for the signal only cross section $\sigma_H$ and the
signal+interference cross section $\sigma_{Hi}$ for $c_{1,2}=\bar
c_{1,2}$ in Table~\ref{restable}.  To facilitate the comparison with
the results of Ref.~\cite{keith}, LO results are computed using NLO
PDFs.  For the signal, the quoted error is obtained by comparing our
soft-collinear approximation to the $N$-soft approximation. For the
background, we also consider the additional uncertainty coming from
independently varying the $c_{1,2}$ coefficients for the first two and the
third generation in the $-5 \bar c_{1,2}<c_{1,2}<5 \bar c_{1,2}$ range. This
leads to an uncertainty of about $6\%$ on the interference predictions
which, combined with the uncertainty of the soft approximation, gives
an overall uncertainty of about $8-9\%$ at NNLO, see
Table~\ref{restable}.  This uncertainty is of same order of magnitude
as the current uncertainties in the Higgs production rate $\sigma_{\rm
  NNLO}$ related to higher-order QCD radiative corrections, PDF and
$\as$ uncertainties etc, see~\cite{hwg}.  We conclude that our
approach to estimate higher order corrections to the signal-background
interference in the Higgs production offers a robust framework and
adequate phenomenological precision.


\begin{table}[t]
\centering
\begin{ruledtabular}
\begin{tabular}{clllclll}
& \multicolumn{3}{c}{$\sqrt s=8$~TeV} && \multicolumn{3}{c}{$\sqrt s=13$~TeV}\\
\cline{2-4}\cline{6-8}
& LO & NLO & NNLO && LO &NLO &NNLO\\
\hline
$\sigma_H$ & 0.379       & 0.83(2)   & 1.07(5)  && 1.55      & 3.29(8)  & 4.2(2) \\
$\sigma_{Hi}$ & 0.427     & 0.93(3)    & 1.20(7) && 1.66      & 3.5(1)   & 4.5(2) \\
$\sigma_H/\sigma_H^{\rm LO}$ & --- & 2.19(5) & 2.8(1) && --- & 2.13(5) & 2.7(1) \\
$\sigma_{Hi}/\sigma_{Hi}^{\rm LO}$ & --- & 2.19(7) & 2.8(2) && --- & 2.12(6) & 2.7(1) \\
\end{tabular}
\end{ruledtabular}
\caption{Same as Table~\ref{restable}, but with Higgs-based cuts on
  the final state. See text for details.}
\label{restable_cut}
\end{table}

\begin{figure*}[t]
\includegraphics[scale=0.7]{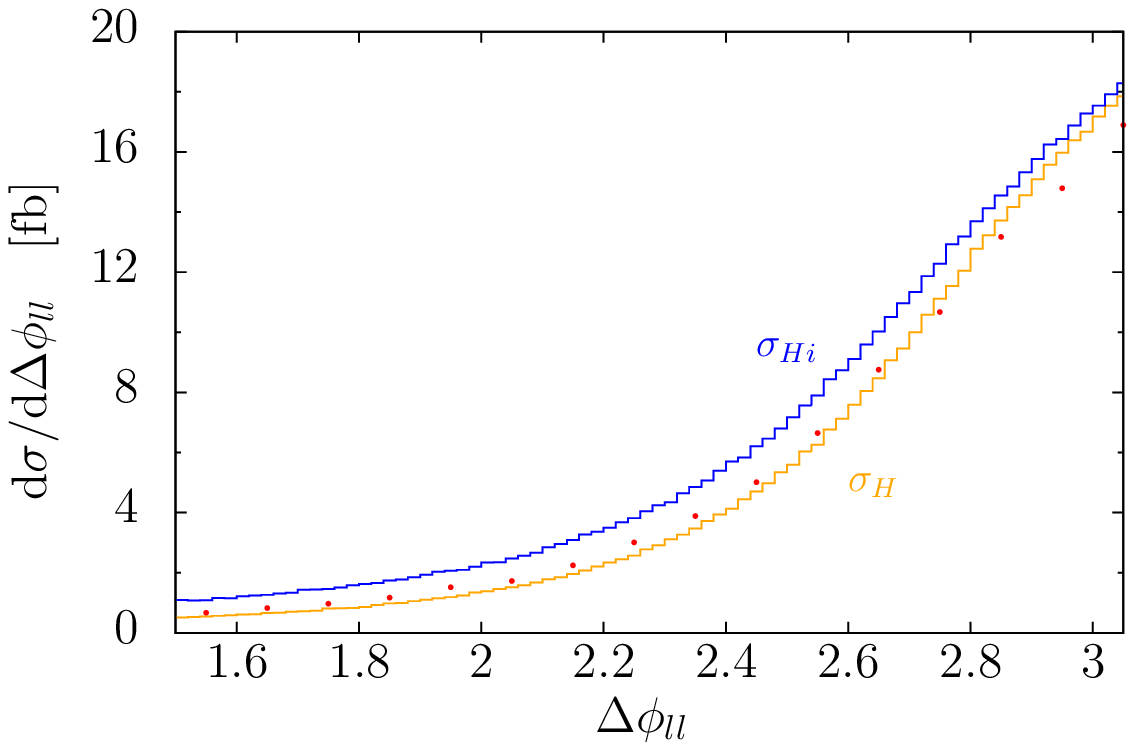}~~~~~~~
\includegraphics[scale=0.7]{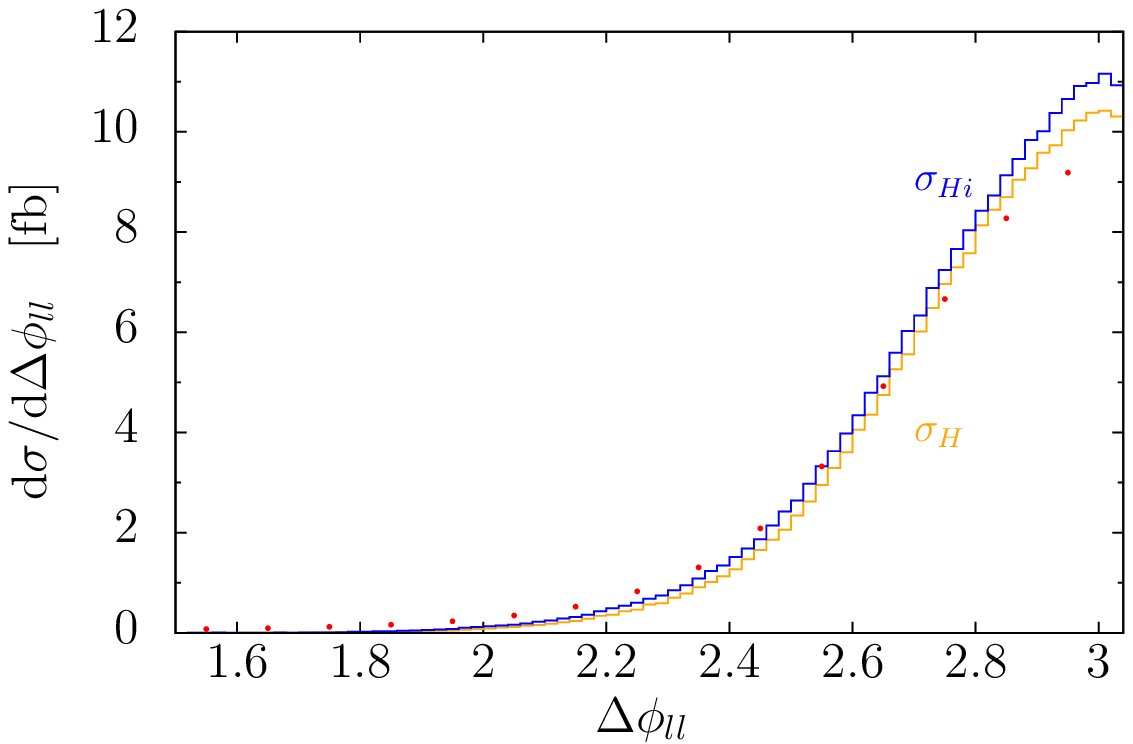}
\caption{Lepton azimuthal distance $\Delta\phi_{ll}$ 
distribution in the fully inclusive case (left pane) and with experimental cuts (right pane) computed with 
the NNLO QCD soft-collinear approximation described in the text. 
Dots show the rescaled MCFM result for the signal ${\rm d}\sigma_{\rm NLO}^{\rm MCFM} \times K_{\rm NNLO}/K_{\rm NLO}$,
where $K_{\rm (N)NLO}$ is the inclusive $K$-factor.}
\label{fig:delphi}
\end{figure*}
\begin{figure*}[t]
\vskip0.4cm
\includegraphics[scale=0.72]{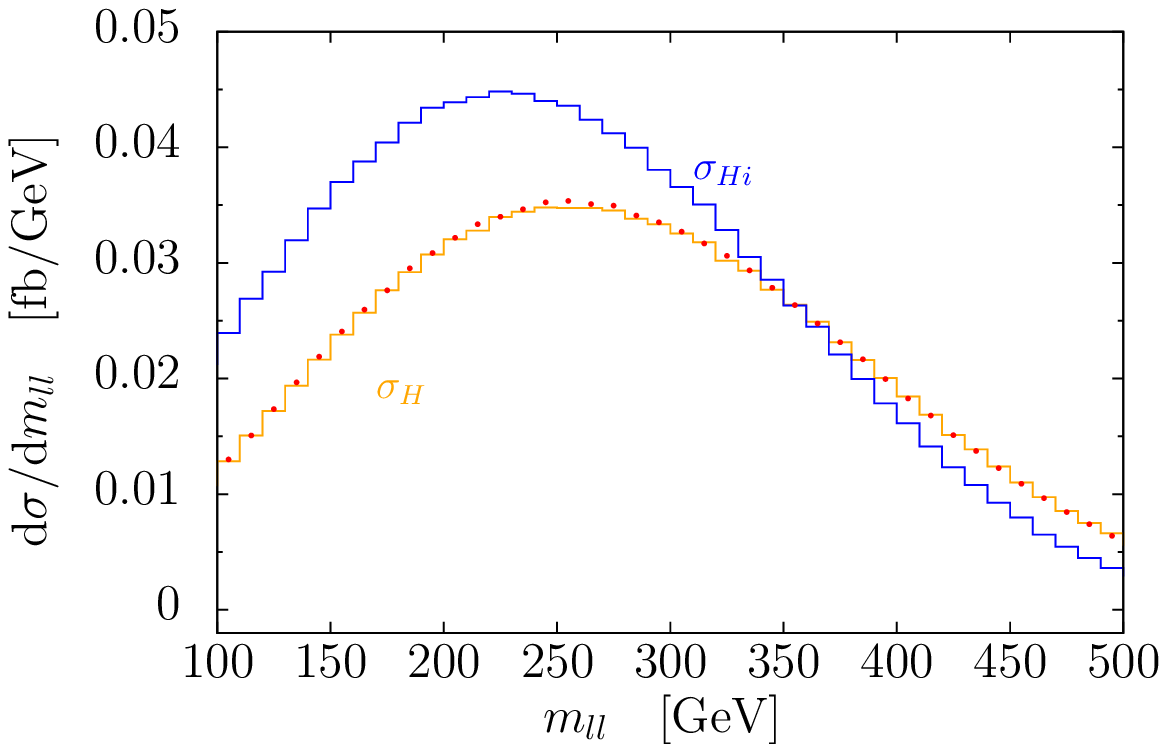}~~~~~~~~
\includegraphics[scale=0.72]{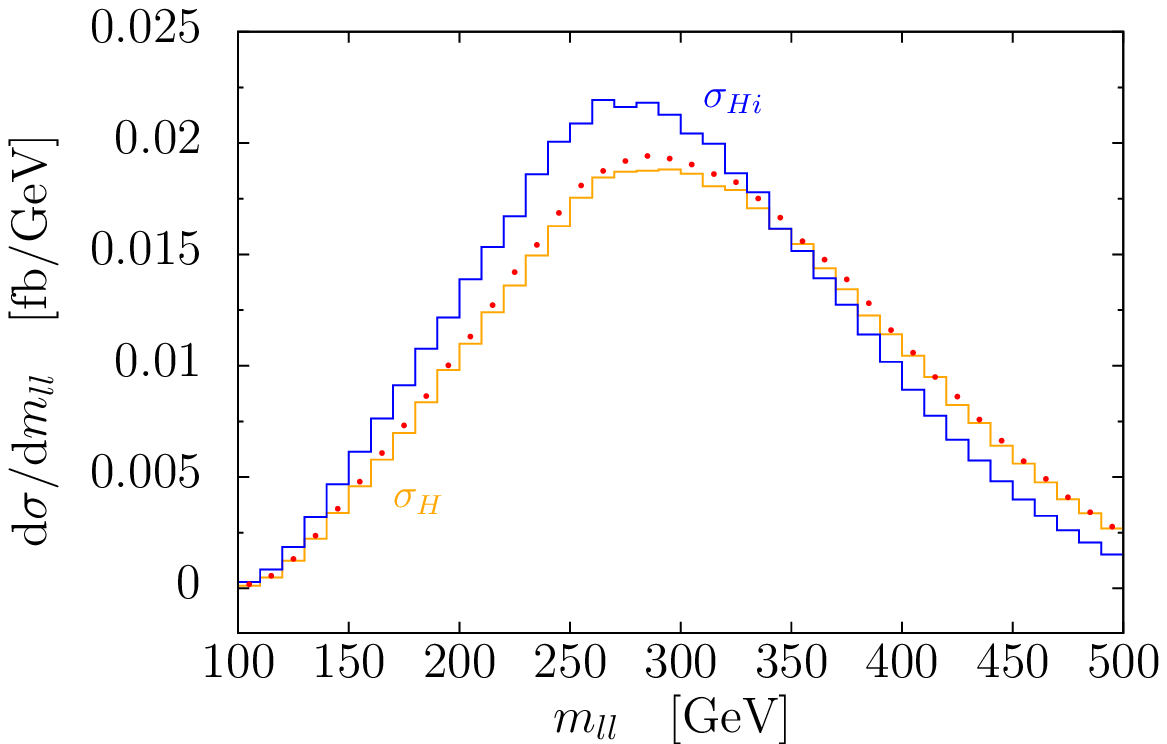}
\caption{Same as Fig.~\ref{fig:delphi}, but for the lepton invariant mass $m_{ll}$ distribution.}
\label{fig:mll}
\end{figure*}
We turn to a discussion of the impact of the interference in a more realistic 
setup,  by imposing selection cuts on  leptons and neutrinos.  
Apart from the standard acceptance cuts on the lepton rapidity $\eta_l$,
lepton transverse momentum $p_t$ and missing energy $\met$,
\be
|\eta_l|<2.5,\qquad p_t > 25~{\rm GeV},\qquad \met > 20~{\rm GeV}
\ee
we impose additional signal-enhancement cuts, 
linearly extrapolating  numerical values given 
in  Ref.~\cite{cms}. To this end,  
we require at least one lepton with $p_t>130$~GeV, and impose the following 
cuts on the lepton invariant mass $m_{ll}$,
azimuthal separation $\Delta\phi_{ll}$ of the two leptons
and transverse mass of the $W^+W^-$ pair $m_\perp$:
\begin{align}
& m_{ll} < 500~{\rm GeV},\qquad
\Delta\phi_{ll} < 3.05,
\nonumber\\
& ~~~~~~~120~{\rm GeV}< m_\perp < m_h.
\end{align}

We note that we have validated the soft-collinear approximation at NLO
QCD against MCFM for the differential distributions, so that we
believe that our results are reliable even when cuts on the final
state are imposed.  We report our results in Tab.~\ref{restable_cut}.
We see that the impact of the interference is mildly (but notably)
reduced when the Higgs-selection cuts are applied to the final state
particles. Note also that radiative corrections to
the interference  are rather similar to  corrections to 
the signal cross section.

We conclude this Section by showing the effect of the interference on
selected kinematic distributions at the $13$~TeV LHC.  In
Fig.~\ref{fig:delphi} we plot the difference of the azimuthal angle
$\Delta\phi_{ll}$ of the two charged leptons with (right pane) and
without (left pane) Higgs-selection cuts. In Fig.~\ref{fig:mll} we do
the same for the invariant mass of the charged leptons $m_{ll}$.  We
plot the NNLO QCD results obtained with our soft-collinear
approximation as described in Sect.~\ref{setup}, using $c_{1,2}=\bar
c_{1,2}$ for the interference case.  We see that the Higgs-selection
cuts reduce the importance of the interference, as already seen in the
total rate.  

An interesting feature of our results is that our approximation reproduces,
to a good accuracy, all the kinematic distributions as obtained with
MCFM.  In particular, all the distributions can be perfectly
reproduced by rescaling the MCFM leading order distributions by the
inclusive NNLO $K$-factor.  For the signal, we also compare our NNLO
approximation against the known NLO distributions, rescaled by the
NNLO/NLO inclusive $K$-factor (also shown in the plots).  Also in this
case, the agreement is excellent; the only exception is the azimuthal
angle distribution where differences are seen at large relative
angles.  This is due to the fact that our soft-collinear approximation
does not reproduce the effects of a hard emission, which modify the
angular distribution.  Note, however, that the azimuthal angle cut
plays an insignificant role in separating the {\it heavy} Higgs
boson from the background so that the impact of this mismatch on
corrections to the interference is minor.

\section{Conclusions}
\label{conc}

We have estimated the impact of QCD
radiative corrections on the signal-background interference in $gg \to
H \to W^+W^-$ process for a heavy Higgs boson.  We constructed a
soft-collinear approximation to higher-order QCD corrections and
verified its validity by comparing it to exact results for $gg \to H$,
including kinematic distributions of the Higgs decay products. We find
that QCD radiative corrections enhance the signal-background
interference by a significant amount which, however, is very similar
to the perturbative QCD enhancement of the signal cross section.

\begin{acknowledgments}
This research is partially supported by US NSF under grants
PHY-1214000. SF and GR are partly supported by a PRIN2010 grant.
Calculations reported in this paper were performed on the Homewood High Performance Cluster of Johns Hopkins University.
\vskip0.2cm
\end{acknowledgments}

\end{document}